\documentstyle{article}

\newcommand{\bundle}[3]{#2\stackrel{#3}{\rightarrow}#1}

\newcommand{\formspace}[4]{#1^{#2}_{#3}#4}
\newcommand{\baseforms}[3]{\formspace{\Lambda}{#1}{#3}{#2}}
\newcommand{\jetforms}[4]{\formspace{\Omega}{#1}{#4#2}{#3}}
\newcommand{\funcforms}[3]{\formspace{\Sigma}{}{#3#1}{#2}}
\newcommand{\funcvecs}[3]{\formspace{\Sigma}{#1}{#3}{#2}}

\newcommand{\parderv}[5]{\frac{\partial^{#2}#1}{\partial#3^{#4}_{#5}}}
\newcommand{\funcderv}[3]{\frac{\delta#1}{\delta#2^{#3}}}

\newcommand{\intprod}{\raisebox{.5mm}{$\:{\scriptscriptstyle\bf\rfloor}\,$}}
\newcommand{\da}{\downarrow}
\newcommand{\ra}{\rightarrow}
\newcommand{\vbicomplex}[8]
{\begin{array}{cccccccccccc}

&&\makebox[0mm]{$0$}&&\makebox[0mm]{$0$}&&&&&&&\\

\vspace{#2}\\

&&\makebox[0mm]{$\da$}&&\makebox[0mm]{$\da$}&&&&&&\\

\vspace{#2}\\

&&\makebox[0mm]{\bf R}&&\makebox[0mm]{\bf
R}&&\makebox[0mm]{$0$}&&\makebox[0mm]{$0$}&&\makebox[0mm]{$0$}&\\

\vspace{#2}\\

&&\makebox[0mm]{$\da$}&&\makebox[0mm]{$\da$}&&\makebox[0mm]{$\da$}&&
\makebox[0mm]{$\da$}&&\makebox[0mm]{$\da$}&\\

\vspace{#2}\\

\makebox[0mm]{$0$} &
\makebox[#1]{$\ra$} &
\makebox[0mm]{$\baseforms{0}{#3}{#7}$} &
\makebox[#1]{$\stackrel{#5^\ast_\infty}{\ra}$} &
\makebox[0mm]{$\jetforms{0}{0}{#5}{#7}$} &
\makebox[#1]{$\stackrel{d_v}{\ra}$} &
\makebox[0mm]{$\jetforms{0}{1}{#5}{#7}$} &
\makebox[#1]{$\cdots$} &
\makebox[0mm]{$\jetforms{0}{s}{#5}{#7}$} &
\makebox[#1]{$\stackrel{d_v}{\ra}$} &
\makebox[0mm]{$\jetforms{0}{s+1}{#5}{#7}$} &
\makebox[#1]{$\cdots$} \\

\vspace{#2}\\

&&
\makebox[0mm]{$\makebox[0mm][r]{$\scriptstyle d$}\da$} &
&
\makebox[0mm]{$\makebox[0mm][r]{$\scriptstyle{#6}$}\da$} &
&
\makebox[0mm]{$\makebox[0mm][r]{$\scriptstyle -{#6}$}\da$} &
&
\makebox[0mm]{$\makebox[0mm][r]{$\scriptstyle (-)^{s}{#6}$}\da$} &
&
\makebox[0mm]{$\makebox[0mm][r]{$\scriptstyle (-)^{s+1}{#6}$}\da$} &
\\

\vspace{#2}\\

\makebox[0mm]{$0$} &
\makebox[#1]{$\ra$} &
\makebox[0mm]{$\baseforms{1}{#3}{#7}$} &
\makebox[#1]{$\stackrel{#5^\ast_\infty}{\ra}$} &
\makebox[0mm]{$\jetforms{1}{0}{#5}{#7}$} &
\makebox[#1]{$\stackrel{d_v}{\ra}$} &
\makebox[0mm]{$\jetforms{1}{1}{#5}{#7}$} &
\makebox[#1]{$\cdots$} &
\makebox[0mm]{$\jetforms{1}{s}{#5}{#7}$} &
\makebox[#1]{$\stackrel{d_v}{\ra}$} &
\makebox[0mm]{$\jetforms{1}{s+1}{#5}{#7}$} &
\makebox[#1]{$\cdots$} \\

\vspace{#8}\\

&&\makebox[0mm]{$\vdots$}&&\makebox[0mm]{$\vdots$}&&\makebox[0mm]{$\vdots$}&&
\makebox[0mm]{$\vdots$}&&\makebox[0mm]{$\vdots$}&\\

\vspace{#8}\\

\makebox[0mm]{$0$} &
\makebox[#1]{$\ra$} &
\makebox[0mm]{$\baseforms{{#4}-1}{#3}{#7}$} &
\makebox[#1]{$\stackrel{#5^\ast_\infty}{\ra}$} &
\makebox[0mm]{$\jetforms{{#4}-1}{0}{#5}{#7}$} &
\makebox[#1]{$\stackrel{d_v}{\ra}$} &
\makebox[0mm]{$\jetforms{{#4}-1}{1}{#5}{#7}$} &
\makebox[#1]{$\cdots$} &
\makebox[0mm]{$\jetforms{{#4}-1}{s}{#5}{#7}$} &
\makebox[#1]{$\stackrel{d_v}{\ra}$} &
\makebox[0mm]{$\jetforms{{#4}-1}{s+1}{#5}{#7}$} &
\makebox[#1]{$\cdots$} \\

\vspace{#2}\\

&&
\makebox[0mm]{$\makebox[0mm][r]{$\scriptstyle (-)^{{#4}-1}d$}\da$} &
&
\makebox[0mm]{$\makebox[0mm][r]{$\scriptstyle (-)^{{#4}-1}{#6}$}\da$} &
&
\makebox[0mm]{$\makebox[0mm][r]{$\scriptstyle (-)^{#4}{#6}$}\da$} &
&
\makebox[0mm]{$\makebox[0mm][r]{$\scriptstyle (-)^{{#4}+s-1}{#6}$}\da$} &
&
\makebox[0mm]{$\makebox[0mm][r]{$\scriptstyle (-)^{{#4}+s}{#6}$}\da$} &
\\

\vspace{#2}\\

\makebox[0mm]{$0$} &
\makebox[#1]{$\ra$} &
\makebox[0mm]{$\baseforms{#4}{#3}{#7}$} &
\makebox[#1]{$\stackrel{#5^\ast_\infty}{\ra}$} &
\makebox[0mm]{$\jetforms{#4}{0}{#5}{#7}$} &
\makebox[#1]{$\stackrel{d_v}{\ra}$} &
\makebox[0mm]{$\jetforms{#4}{1}{#5}{#7}$} &
\makebox[#1]{$\cdots$} &
\makebox[0mm]{$\jetforms{#4}{s}{#5}{#7}$} &
\makebox[#1]{$\stackrel{d_v}{\ra}$} &
\makebox[0mm]{$\jetforms{#4}{s+1}{#5}{#7}$} &
\makebox[#1]{$\cdots$} \\

\vspace{#2}\\

&&
\makebox[0mm]{$\da$} &
&
\makebox[0mm]{$\makebox[0mm][r]{$\int$}\da$} &
&
\makebox[0mm]{$\makebox[0mm][r]{$\int$}\da$} &
&
\makebox[0mm]{$\makebox[0mm][r]{$\int$}\da$} &
&
\makebox[0mm]{$\makebox[0mm][r]{$\int$}\da$} &
\\

\vspace{#2}\\

&&
\makebox[0mm]{$0$} &
\makebox[#1]{$\ra$} &
\makebox[0mm]{$\funcforms{0}{#5}{#7}$} &
\makebox[#1]{$\stackrel{\delta}{\ra}$} &
\makebox[0mm]{$\funcforms{1}{#5}{#7}$} &
\makebox[#1]{$\cdots$} &
\makebox[0mm]{$\funcforms{s}{#5}{#7}$} &
\makebox[#1]{$\stackrel{\delta}{\ra}$} &
\makebox[0mm]{$\funcforms{s+1}{#5}{#7}$} &
\makebox[#1]{$\cdots$} \\

\vspace{#2}\\

&&&&\makebox[0mm]{$\da$}&&\makebox[0mm]{$\da$}&&\makebox[0mm]{$\da$}&&
\makebox[0mm]{$\da$}&\\

\vspace{#2}\\

&&&&\makebox[0mm]{$0$}&&\makebox[0mm]{$0$}&&\makebox[0mm]{$0$}&&
\makebox[0mm]{$0$}&\\

\end{array}}

\begin{document}

\title{Jet Bundles in Quantum Field Theory:\\The BRST-BV method}
\author{Paul McCloud\\Department of Mathematics, Kings College London}
\maketitle

\begin{abstract}
The geometric interpretation of the Batalin-Vilkovisky antibracket as the
Schouten bracket of functional multivectors is examined in detail. The
identification is achieved by the process of repeated contraction of even
functional multivectors with fermionic functional 1-forms. The classical master
equation may then be considered as a generalisation of the Jacobi identity for
Poisson brackets, and the cohomology of a nilpotent even functional multivector
is identified with the BRST cohomology. As an example, the BRST-BV formulation
of gauge fixing in theories with gauge symmetries is reformulated in the jet
bundle formalism.
\end{abstract}

\pagebreak

\section{Introduction}

The jet bundles to a manifold provide a geometric formulation of the theory of
partial differential equations. Coordinates on the jet bundle are given by the
independent variables, the dependent variables, and the derivatives of the
dependent variables with respect to the independent variables. PDEs are
identified with submanifolds of the jet bundle, (infinitesimal) symmetries of
the PDEs then occur naturally as tangent vector fields on the jet bundle with
flows preserving the submanifolds, \cite[chapter 3]{jb:Olver86}. Similarly,
symmetries of the action in the Lagrangian formulation of classical field
theory occur as evolutionary vector fields on the jet bundle,
\cite[chapters 4 and 5]{jb:Saunders89,jb:Olver86}.

The location of symmetries in the classical field theory reduces to a geometric
procedure on the jet bundle. The question remains: can the symmetries be
maintained in the quantum field theory? The BRST-BV approach to this question
is to consider an extended classical action, dependent on the bosonic fields
(only bosonic fields are considered in this article, though the extension to
include fermionic fields is clear) and also the external fermionic antifields,
\cite{bv:BatVil81,bv:BatVil83}. When the extended action satisfies the
classical master equation, the Green's functions for symmetric sources are
independent of the antifields up to anomalous terms. One consequence of this is
that, in the absence of anomaly terms, the quantisation procedure may be
formulated so that the symmetry is preserved in the quantum effective action.
Another consequence is the formulation of gauge fixed quantum field theories.

It is natural to consider the jet formulation of the BRST-BV method. The series
expansion of a local functional of the fermionic antifield corresponds to a sum
of repeatedly contracted functional multivectors on the jet bundle. Moreover,
the antifield formalism is valid only on vector bundles, whilst the jet bundle
formulation naturally generalises to any bundle, in particular to sigma models.

The first section of this article constitutes a brief introduction to the
theory of jet bundles, a more complete review may be found in
\cite{jb:Olver86,jb:Saunders89}. The jet bundle analysis of Lagrangian
dynamical systems has the advantage of explicitly maintaining locality,
reducing a variational problem to a system of PDE's. Functional multivectors
are introduced and their basic properties considered. Functional bivectors have
been employed in Hamiltonian theory, see
\cite{jb:Olver80,jb:Magri78,jb:GelDor79,jb:GelDor80,jb:GelDor81} and chapters 6
and 7 of \cite{jb:Olver86}. The cohomology of nilpotent bivectors is analysed
in the finite dimensional case in \cite{jb:Lichnerowicz77,jb:Lichnerowicz82}.
The generalisation to arbitrary even functional multivectors considered in this
article does not present any extra difficulties. Local bosonic functionals of a
fermionic field may then be constructed as repeatedly contracted even
functional multivectors.

The next section developes the classical and quantum theories in the jet
formalism, in a very general setting, and compares this with the BRST-BV
field-antifield formalism,
\cite{bv:BatVil81,bv:BatVil83,bv:BatTyu90,bv:BatTyu92,bv:FerHenPuc92,bv:GreHen92,bv:Henneaux92}.
The jet formulation of the quantum field theory is valid whenever a definition
of the functional integral exists. In particular nonlinear bundles, where no
global definition of the antifields exists, present no difficulties in the jet
approach. Instead of considering local functionals of the bosonic field and its
fermionic antifield, an even functional multivector is repeatedly contracted
with a fermionic functional 1-form. This fermionic functional 1-form
generalises both the antifields and the gauge fixing fermion in one object. The
classical master equation then becomes a nilpotency condition on the even
functional multivector.

The anomalies of the system are interpreted as odd functional multivectors. The
Lam action principle \cite{qft:Lam72,qft:ClaLow76} may be constructed for the
quantum effective action. This is then used to generate the anomaly consistency
condition, \cite{qft:PigRou81,bv:DiaTroNiePro89,bv:TroNiePro90,bv:HowLinWhi90}.
In the jet formalism, the consistency condition implies that the anomalies
define elements of the odd cohomology group of the nilpotent even functional
multivector.

The final section analyses the problem of gauge fixing in the jet formalism.
The gauge symmetries naturally define a functional bivector that in turn may be
used to generate the symmetries, and the algebra the symmetries form defines a
nilpotent functional bivector. A combination of these functional bivectors
solves the classical master equation and may be used to extend the classical
action. It is observed that the fermionic functional 1-form contains both the
gauge fixing fermion and the ghost field in one object, and in this way the jet
formalism reproduces the BRST-BV results,
\cite{bv:BatVil81,bv:BatVil83,qft:PigRou81,bv:TroNiePro90}.

A future paper will consider the use of jet bundles in the analysis of chiral
and W-symmetry algebras and the anomalies in these algebras. This requires an
extension of the jet bundle theory to bundles with a chiral decomposition of
the tangent bundle. W-symmetries are discovered, under special conditions, in
arbitrary dimensions, and the nature of the full Lie algebra generated from the
W-algebra is revealed. The classical master equation is solved, thus defining a
cohomology problem for the calculation of anomalies in chiral and W-symmetry
algebras.

\section{Introduction to jet bundles and functional multivectors}

\subsection{Jet bundles}

The notion of the tangent vector to a manifold arises by considering the first
order contact of curves in the manifold. The theory of jets generalises this in
two ways: the contact of higher dimensional curves, i.e.\ submanifolds, and
contact at arbitrarily high orders are considered. This leads to the definition
of the jet bundles of a manifold. The jet bundles form the natural geometric
arena for the study of partial differential equations, and in particular the
equations arising from the Lagrangian approach to field theory.

The jet bundles to a manifold, even a trivial manifold, have non-trivial
geometry. For a fibre bundle, a restricted jet bundle may be defined by
restricting attention to submanifolds occuring as the images of sections of the
bundle. This simplification neglects the possibility of singular and
multi-valued solutions of PDEs, accommodated in the full jet bundle. However,
it is sufficient for the purposes of this article.

Consider a fibre bundle $\bundle{M}{B}{\pi}$ over the compact, connected, and
orientable manifold without boundary $M$. Denote the space of smooth sections
$\phi :M\rightarrow B$ of $\bundle{M}{B}{\pi}$ by $\Gamma(\bundle{M}{B}{\pi})$
or $\Gamma(\pi)$, and the space of local sections near $x_0\in M$ by
$\Gamma_{x_0}(\pi)$. In terms of the local adapted coordinates
$(x^i,\phi^\alpha)$ on $B$, the two local sections
$\phi_1,\phi_2\in\Gamma_{x_0}(\pi)$ are equivalent under the relation of
infinite order contact at $x_0$ if:
\begin{displaymath}
\left.\parderv{(\phi^\alpha\circ\phi_1)}{|I|}{x}{I}{}\right|_{x_0}=\left.
\parderv{(\phi^\alpha\circ\phi_2)}{|I|}{x}{I}{}\right|_{x_0}
\end{displaymath}
for all multi-indices $I=(n_1,\ldots,n_m)$, where $m={\rm dim}M$. The jet space
$J^\infty_{x_0}\pi$ is the space $\Gamma_{x_0}(\pi)$ modulo this relation, and
the collection $J^\infty\pi=\bigcup_{x\in M}J^\infty_x\pi$ constitutes the jet
bundle over $M$, $\bundle{M}{J^\infty\pi}{\pi_\infty}$. Local coordinates on
the jet bundle are given by:
\begin{displaymath}
\phi^\alpha_I([\phi]_{x_0})=\left.\parderv{(\phi^\alpha\circ\phi)}{|I|}{x}{I}{}
\right|_{x_0}
\end{displaymath}

Any section $\phi\in\Gamma(\pi)$ may be prolonged to a section $j^\infty\phi$
of the jet bundle via $j^\infty\phi(x)=[\phi]_x$. A function $f$ on
$J^\infty\pi$ can then be evaluated on sections of $\bundle{M}{B}{\pi}$ as a
function on the base space. The total derivatives $d_i$, tangent vectors on
$J^\infty\pi$, are defined by the expression:
\begin{displaymath}
(d_if)\circ j^\infty\phi=\partial_i(f\circ j^\infty\phi)
\end{displaymath}
In terms of local coordinates:
\begin{displaymath}
d_i=\partial_i+\phi^\alpha_{Ii}\parderv{}{}{\phi}{\alpha}{I}
\end{displaymath}

The total derivatives determine a horizontal subspace to the tangent bundle of
$J^\infty\pi$. Since the total derivatives commute, $[d_i,d_j]=0$, this defines
a canonical integrable connection on the jet bundle:
\begin{displaymath}
VJ^\infty\pi=<\parderv{}{}{\phi}{\alpha}{I}>,\;HJ^\infty\pi=<d_i>
\end{displaymath}

\subsection{Variational bicomplex}

The decomposition of the tangent bundle induces a decomposition of the spaces
of differential forms on the jet space. Denote the space of forms with $r$
horizontal and $s$ vertical factors by $\jetforms{r}{s}{\pi}{}$. The exterior
derivative decomposes, $d=d_h+d_v$, where:
\begin{displaymath}
d_h=dx^i\,d_i,\;d_v=(d\phi^\alpha_I-\phi^\alpha_{Ii}\,dx^i)
\parderv{}{}{\phi}{\alpha}{I}
\end{displaymath}

The properties of the horizontal and vertical exterior derivatives are
summarised in the variational bicomplex,
Figure~\ref{Fig: Variational bicomplex}.
\begin{figure}
$\vbicomplex{13mm}{-1mm}{M}{m}{\pi}{d_h}{}{-2.2mm}$
\caption{The variational bicomplex}
\label{Fig: Variational bicomplex}
\end{figure}
This diagram commutes, and any row or column forms a complex. Furthermore, it
can be demonstrated that the variational bicomplex is locally exact---on
topologically trivial spaces the cohomology of the variational bicomplex is
trivial.

The bicomplex is closed on the first column by the de Rham complex of the base
space, and on the bottom row by the variational complex:
\begin{displaymath}
\begin{array}{cccccccccc}
\makebox[0mm]{$0$} &
\makebox[13mm]{$\ra$} &
\makebox[0mm]{$\funcforms{0}{\pi}{}$} &
\makebox[13mm]{$\stackrel{\delta}{\ra}$} &
\makebox[0mm]{$\funcforms{1}{\pi}{}$} &
\makebox[13mm]{$\cdots$} &
\makebox[0mm]{$\funcforms{s}{\pi}{}$} &
\makebox[13mm]{$\stackrel{\delta}{\ra}$} &
\makebox[0mm]{$\funcforms{s+1}{\pi}{}$} &
\makebox[13mm]{$\cdots$} \\
\end{array}
\end{displaymath}
The functional $s$-forms $\funcforms{s}{\pi}{}$ are given by the quotient:
\begin{displaymath}
\funcforms{s}{\pi}{}=\jetforms{m}{s}{\pi}{}/d_h\jetforms{m-1}{s}{\pi}{}
\end{displaymath}
The projection $\jetforms{m}{s}{\pi}{}\ra\funcforms{s}{\pi}{}$ is denoted by
the integral sign $\int$, and the variational derivative $\delta$ is induced by
the vertical exterior derivative:
\begin{displaymath}
\delta\int\omega=\int d_v\omega
\end{displaymath}
Define the variational cohomology groups:
\begin{displaymath}
\formspace{H}{}{s}{(\pi,\delta)}={\rm Ker}\,\delta\cap\funcforms{s}{\pi}{}/
{\rm Im}\,\delta\cap\funcforms{s}{\pi}{}
\end{displaymath}

The vertical vector fields may be contracted with forms on the jet space,
reducing the vertical degree by one. It is of interest to consider when it is
possible to contract vertical vector fields with functional forms in an
unambiguous manner.
Take $\eta=\eta^i\wedge d_i\intprod\Omega\in\jetforms{m-1}{s}{\pi}{}$, with
$\eta^i\in\jetforms{0}{s}{\pi}{}$, and $v$ a vertical vector field. Then:
\begin{displaymath}
v\intprod d_h\,\eta+d_h(v\intprod\eta)=(-)^s[v,d_i]\intprod\eta^i\wedge\Omega
\end{displaymath}
In particular, when $[v,d_i]=0$ then $v\intprod d_h\,\eta=-d_h(v\intprod\eta)$.
Vertical vector fields satisfying $[v,d_i]=0$ are evolutionary vector fields,
and take the form:
\begin{displaymath}
v=(d_I\,v^\alpha)\parderv{}{}{\phi}{\alpha}{I}
\end{displaymath}
for some $v^\alpha$. Thus evolutionary vector fields may be contracted with
functional forms, $v\intprod\int\omega=\int v\intprod\omega$.

The representative for a functional form is only defined up to a total
horizontal derivative. Integration by parts may thus be performed under the
integral sign. In particular, this may be employed to reduce functional forms
to canonical forms. Take a volume form $\Omega=\rho\,d^mx$ on $M$. Using
integration by parts, functional 1-forms may be expressed in the canonical form
$\int\omega_\alpha\,d\phi^\alpha\wedge\Omega$. In particular:
\begin{displaymath}
\delta\int{\cal L}\Omega=\int E_\alpha({\cal L})\,d\phi^\alpha\wedge\Omega
\end{displaymath}
where $E_\alpha$ is the Euler operator:
\begin{displaymath}
E_\alpha=(-)^{|I|}\rho^{-1}d_I\rho\parderv{}{}{\phi}{\alpha}{I}
\end{displaymath}
Thus $\delta\int{\cal L}\Omega=0$ if and only if $E_\alpha({\cal L})=0$.

\subsection{Evaluation and expansion of functionals}

Many of the properties of the variational cohomology have parallels in the
finite dimensional de Rham cohomology, with functional forms replaced by
differential forms, and evolutionary vector fields replaced with tangent vector
fields. In fact (at least locally) functional forms may be identified with
forms on the space of sections $\Gamma(\pi)$, and evolutionary vector fields
with tangent vector fields on $\Gamma(\pi)$. The Taylor series expansion of
functional 0-forms may thus be considered analogously to the expansion of
functions on manifolds.

Horizontal forms on the jet bundle may be evaluated on sections as forms on the
base space. By definition of the total derivatives,
$d_h\eta\circ j^\infty\phi=d(\eta\circ j^\infty\phi)$ for
$\eta\in\jetforms{m-1}{0}{\pi}{}$. Since $M$ is compact and boundary-less, then
by Stokes' theorem functional 0-forms may be evaluated unambiguously on
sections---for $S=\int{\cal L}\Omega\in\funcforms{0}{\pi}{}$ then:
\begin{displaymath}
S[\phi]=\int_M{\cal L}\circ j^\infty\phi\Omega
\end{displaymath}
The functional 0-forms constitute the local functionals in
$\Lambda^0\Gamma(\pi)$.

The tangent vectors on $\Gamma(\pi)$ are transformations
$\delta\phi^\alpha=v^\alpha$, and are identified with evolutionary vector
fields. The requirement that evolutionary vector fields commute with the total
derivatives thus corresponds to the rule $[\delta,d_i]=0$ for transformations.
In particular, for $\int\omega\in\funcforms{s}{\pi}{}$ and $v_1,\ldots,v_s$
evolutionary vector fields, then $(v_1,\ldots,v_s)\intprod\int\omega$ is a
local functional. Thus the functional $s$-forms constitute the local $s$-forms
on $\Gamma(\pi)$.

The transformation $\delta\phi^\alpha=v^\alpha$ may be exponentiated to give a
flow $e^{\lambda v}$ on $\Gamma(\pi)$. The variational derivative of a
functional $S$ is defined by:
\begin{displaymath}
\frac{d}{d\lambda}S[e^{\lambda v}\phi_0]=\int
v^\alpha\funcderv{S}{\phi}{\alpha}\Omega[e^{\lambda v}\phi_0]
\end{displaymath}
For the local functional $S=\int{\cal L}\Omega$, then
$\delta S/\delta\phi^\alpha=E_\alpha({\cal L})$. The Taylor series expansion of
$S$ is thus:
\begin{eqnarray*}
S[e^{\lambda v}\phi_0] &
= &
\sum^\infty_{n=0}\frac{\lambda^n}{n!}(v\intprod\delta)^nS[\phi_0]\\
& = &
S[\phi_0]+\sum^\infty_{n=0}\frac{\lambda^{n+1}}{(n+1)!}
v\intprod v^n(\delta S)[\phi_0]
\end{eqnarray*}
When $\bundle{M}{B}{\pi}$ is a vector bundle, take the transformation $v$ to be
a translation. Then the expansion above for a local functional gives:
\begin{displaymath}
S[\phi]=S[0]+\sum^\infty_{n=0}\frac{1}{(n+1)!}\int
\phi^\alpha\phi^{\beta_1}_{I_1}\ldots\phi^{\beta_n}_{I_n}
{S_\alpha}^{I_1\ldots I_n}_{\beta_1\ldots\beta_n}\Omega
\end{displaymath}
with coefficient functions:
\begin{displaymath}
{S_\alpha}^{I_1\ldots
I_n}_{\beta_1\ldots\beta_n}=\frac{\partial^n}{\partial\phi^{\beta_1}_{I_1}
\ldots\partial\phi^{\beta_n}_{I_n}}\left.\left(\funcderv{S}{\phi}{\alpha}
\right)\right|_{\phi=0}
\end{displaymath}
The expansion of the classical action defines the propagators and vertices of
the perturbative quantum field theory, in which case the coefficient functions
above would be renormalised in order to maintain finiteness. The above
expansion may also be employed for chiral local functionals to give a basis for
the full Lie algebra generated from a W-algebra. This will be explained in a
future paper.

\subsection{Functional multivectors}

The parallel between variational cohomology and de Rham cohomology fails when
considering the wedge product of differential forms. The wedge product of two
local functional forms is bilocal, and hence the space of functional forms does
not close under the product---the functional forms do not possess a wedge
product structure. The definition of a dual space to the functional forms is
thus ambiguous. The functional $s$-forms $\funcforms{s}{\pi}{}$ define
alternating, $s$-linear maps on the evolutionary vector fields, valued in the
space of local functionals $\funcforms{0}{\pi}{}$. Define the dual space
$\funcvecs{s}{\pi}{}$ of functional $s$-vectors to be the space of alternating,
$s$-linear maps $(\funcforms{1}{\pi}{})^s\ra\funcforms{0}{\pi}{}$. For
completion, define also the functional 0-vectors $\funcvecs{0}{\pi}{}$ to be
the local functionals $\funcforms{0}{\pi}{}$.

Since the canonical form for $\omega\in\funcforms{1}{\pi}{}$ is
$\omega=\int\omega_\alpha\,d\phi^\alpha\wedge\Omega$, the functional $s$-vector
$v\in\funcvecs{s}{\pi}{}$ may be written:
\begin{displaymath}
v=\int\Omega_1\ldots\int\Omega_s\,v^{\alpha_1\ldots\alpha_s}
\frac{\partial}{\partial\phi^{\alpha_1}_1}\wedge\ldots\wedge
\frac{\partial}{\partial\phi^{\alpha_s}_s}
\end{displaymath}
with the obvious definition of contraction with functional 1-forms. The result
of contraction must be a local functional. The coefficient functions must thus
be products of $\delta$-functions, of the form:
\begin{displaymath}
v^{\alpha_1\ldots\alpha_s}=\int(-)^{|I_1|+\cdots+|I_s|}
v^{\alpha_1I_1\ldots\alpha_sI_s}\rho^{-1}_1d^{(1)}_{I_1}(\rho_1\delta_x(x_1))
\ldots\rho^{-1}_sd^{(s)}_{I_s}(\rho_s\delta_x(x_s))\Omega
\end{displaymath}
where the $\delta$-function satisfies $f(x)=\int f(y)\delta_x(y)\Omega_y$.
Performing the integrals gives the general form of the functional $s$-vector:
\begin{displaymath}
v=\int\Omega\,v^{\alpha_1I_1\ldots\alpha_sI_s}d_{I_1}
\parderv{}{}{\phi}{\alpha_1}{}\wedge\ldots
\wedge d_{I_s}\parderv{}{}{\phi}{\alpha_s}{}
\end{displaymath}
so that for the functional 1-form
$\omega=\int\omega_\alpha\,d\phi^\alpha\wedge\Omega$:
\begin{displaymath}
v\intprod\omega=\sum^s_{r=1}(-)^{r-1}\int\Omega\,
v^{\alpha_1I_1\ldots\alpha_sI_s}(d_{I_r}\omega^{\alpha_r})d_{I_1}
\parderv{}{}{\phi}{\alpha_1}{}\wedge\ldots
\wedge d_{I_s}\parderv{}{}{\phi}{\alpha_s}{}
\end{displaymath}
where the wedge product in the above expression omits the $r$th factor.

The coordinate representation above for the functional $s$-vector $v$ is not
unique. As for functional forms, integration by parts may be performed under
the integral sign. Functional multivectors may thus be reduced to canonical
forms---the functional vectors have the canonical form:
\begin{displaymath}
\int\Omega\,v^\alpha\parderv{}{}{\phi}{\alpha}{}
\end{displaymath}
and are identified with the evolutionary vector fields, whilst the functional
bivectors have the canonical form:
\begin{displaymath}
\frac{1}{2}\int\Omega\,\parderv{}{}{\phi}{\alpha}{}\wedge
D^{\alpha\beta}\parderv{}{}{\phi}{\beta}{}
\end{displaymath}
for some skew adjoint operator $D^{\alpha\beta}$.

Using the coordinate expression for the functional $s$-vector $v$, the
variational derivative $\delta v$ may be defined:
\begin{displaymath}
\delta v=\int d\phi^\beta_J\wedge\Omega\,
\parderv{v^{\alpha_1I_1\ldots\alpha_sI_s}}{}{\phi}{\beta}{J}d_{I_1}
\parderv{}{}{\phi}{\alpha_1}{}\wedge\ldots
\wedge d_{I_s}\parderv{}{}{\phi}{\alpha_s}{}
\end{displaymath}
This variational derivative does not have a coordinate independent meaning
(without the introduction of a connection on the space of sections
$\Gamma(\pi)$), and cannot be extended to a global definition. However, the
combination:
\begin{displaymath}
[v_1,v_2]=v_1\intprod\delta v_2+(-)^{s_1s_2}v_2\intprod\delta
v_1\in\funcvecs{s_1+s_2-1}{\pi}{}
\end{displaymath}
for the functional multivectors $v_{1,2}\in\funcvecs{s_1,s_2}{\pi}{}$ is
globally defined, and defines the Schouten bracket of $v_1$ and $v_2$. The
bracket is `antisymmetric':
\begin{displaymath}
[v_1,v_2]=(-)^{s_1s_2}[v_2,v_1]
\end{displaymath}
and satisfies the `Jacobi identity':
\begin{displaymath}
(-)^{s_1(s_3+1)}[v_1,[v_2,v_3]]+{\rm cyclic\;permutations}=0
\end{displaymath}
for $v_{1,2,3}\in\funcvecs{s_1,s_2,s_3}{\pi}{}$. Furthermore, if
$\omega\in\funcforms{1}{\pi}{}$ is a closed functional 1-form,
$\delta\omega=0$, then the bracket satisfies the closed form identity:
\begin{displaymath}
[v_1,v_2]\intprod\omega+[v_1\intprod\omega,v_2]+
(-)^{s_1}[v_1,v_2\intprod\omega]=0
\end{displaymath}

\subsection{Multivector cohomology}

The Schouten bracket of the evolutionary vector field $v$ with the local
functional $S$ is simply the variation of $S$, $vS\equiv v\intprod\delta S$,
whilst the Schouten bracket of two evolutionary vector fields is the familiar
vector field commutator. Consider the interpretation of the bracket for general
functional multivectors.

For the even functional multivector
$\Gamma\in\bigoplus^\infty_{r=0}\funcvecs{2r}{\pi}{}$, the Jacobi identity for
the Schouten bracket gives:
\begin{displaymath}
[\Gamma,[\Gamma,v]]=-\frac{1}{2}[[\Gamma,\Gamma],v]
\end{displaymath}
for any functional multivector $v$. When $\Gamma$ is nilpotent,
$[\Gamma,\Gamma]=0$, the commutator action $\Gamma:v\mapsto[\Gamma,v]$ is
nilpotent. In this case, $\Gamma$ defines a complex:
\begin{displaymath}
\bigoplus^\infty_{r=0}\funcvecs{2r}{\pi}{}
\begin{array}{c}
{\scriptstyle \Gamma}\\\rightleftharpoons\\{\scriptstyle \Gamma}
\end{array}
\bigoplus^\infty_{r=0}\funcvecs{2r+1}{\pi}{}
\end{displaymath}
Define the even and odd $\Gamma$-cohomology groups:
\begin{eqnarray*}
H^{\rm even}(\pi,\Gamma)&=
&{\rm Ker}\,\Gamma\cap\bigoplus^\infty_{r=0}\funcvecs{2r}{\pi}{}/
{\rm Im}\,\Gamma\cap\bigoplus^\infty_{r=0}\funcvecs{2r}{\pi}{}\\
H^{\rm odd}(\pi,\Gamma)&=
&{\rm Ker}\,\Gamma\cap\bigoplus^\infty_{r=0}\funcvecs{2r+1}{\pi}{}/
{\rm Im}\,\Gamma\cap\bigoplus^\infty_{r=0}\funcvecs{2r+1}{\pi}{}
\end{eqnarray*}
Thus nilpotent even functional multivectors define cohomology operators on the
functional multivectors, via the Schouten bracket action.

Functional bivectors can be used to construct Poisson brackets on the space of
local functionals. Take $\Gamma$ to be a functional bivector, with skew adjoint
operator $D^{\alpha\beta}$. The Poisson bracket on $\funcforms{0}{\pi}{}$
determined by $\Gamma$ is:
\begin{displaymath}
\{S,T\}=\Gamma\intprod\delta S\intprod\delta T=\int
\funcderv{S}{\phi}{\alpha} D^{\alpha\beta}\funcderv{T}{\phi}{\beta}\Omega
\end{displaymath}
The nilpotency condition $[\Gamma,\Gamma]=0$ and the Jacobi identity for the
Poisson bracket are equivalent:
\begin{displaymath}
\{R,\{S,T\}\}+{\rm
cyclic\;permutations}=-\frac{1}{2}[\Gamma,\Gamma]\intprod\delta R\intprod
\delta S\intprod\delta T
\end{displaymath}
i.e.\ the Poisson bracket satisfies the Jacobi identity if and only if $\Gamma$
is nilpotent. In this case, the commutator action of $\Gamma$ gives a complex:
\begin{displaymath}
\begin{array}{cccccccccc}
\makebox[0mm]{$0$} &
\makebox[13mm]{$\ra$} &
\makebox[0mm]{$\funcvecs{0}{\pi}{}$} &
\makebox[13mm]{$\stackrel{\Gamma}{\ra}$} &
\makebox[0mm]{$\funcvecs{1}{\pi}{}$} &
\makebox[13mm]{$\cdots$} &
\makebox[0mm]{$\funcvecs{s}{\pi}{}$} &
\makebox[13mm]{$\stackrel{\Gamma}{\ra}$} &
\makebox[0mm]{$\funcvecs{s+1}{\pi}{}$} &
\makebox[13mm]{$\cdots$} \\
\end{array}
\end{displaymath}
with cohomology groups:
\begin{displaymath}
\formspace{H}{s}{}{(\pi,\Gamma)}={\rm Ker}\,\Gamma\cap\funcvecs{s}{\pi}{}/
{\rm Im}\,\Gamma\cap\funcvecs{s}{\pi}{}
\end{displaymath}
The cohomology groups may be used to analyse the Poisson bracket, for example
the group $H^0(\pi,\Gamma)$ is identified with the space of distinguished local
functionals, whose Poisson bracket with any other local functional vanishes.

\subsection{Contraction with fermionic functional 1-forms}

The functional multivector, when contracted, is local and antisymmetric in its
arguments. This suggests that they may be used to provide an expansion for
local functionals of fermionic fields. Effectively, the functional multivectors
are identified with the local functionals on a superspace. In fact this is only
possible in the case of a linear bundle, but the idea of repeated contraction
with fermionic functional 1-forms is valid in nonlinear bundles, and provides a
natural generalisation of the notion of the fermionic antifield.

Let $\eta=\int\eta_\alpha\,d\phi^\alpha\wedge\Omega$ be a fermionic functional
1-form. Then $\eta$ may be repeatedly contracted with a functional multivector.
For $v=\sum^\infty_{s=0}v^s$, $v^s\in\funcvecs{s}{\pi}{}$, define:
\begin{displaymath}
v^r_{\eta}=\sum^\infty_{s=r}\frac{1}{(s-r)!}\,v^s\intprod\eta
\stackrel{s-r}{\intprod\cdots\intprod}\eta\in\funcvecs{r}{\pi}{}
\end{displaymath}
The cases $r=0$ and $r=1$ are of particular importance. For $r=0$,
$v_\eta\equiv v^0_{\eta}$ is a local functional, bosonic if $v$ is even,
fermionic if $v$ is odd. For $r=1$, $\hat{v}_\eta\equiv v^1_{\eta}$ is an
evolutionary vector field.

If $\chi$ is another fermionic functional 1-form, the repeated contractions
with the forms $\eta$, $\chi$, and $\eta+\chi$ may all be defined. Separating
out the contractions demonstrates that they are related:
\begin{displaymath}
v^r_{\eta+\chi}=\sum^\infty_{s=r}(v^s_\eta)^r_{\chi}
\end{displaymath}
If $\chi$ is a perturbation of $\eta$, then to lowest order in $\chi$:
\begin{displaymath}
v^r_{\eta+\chi}-v^r_{\eta}=v^{r+1}_\eta\intprod\chi
\end{displaymath}
In the case $r=0$:
\begin{displaymath}
v_{\eta+\chi}-v_\eta=\hat{v}_\eta\intprod\chi
\end{displaymath}
In the linear case this expression would define the variational derivative of
$v_\eta$ with respect to the fermionic antifield $\eta$. Thus in the general
nonlinear case this is replaced with the evolutionary vector field
$\hat{v}_\eta$.

When the fermionic functional 1-form $\eta$ is closed, $\delta\eta=0$,  the
closed form identity can be used to take the contractions into the commutator,
resulting in an expression of the form:
\begin{displaymath}
[v_1,v_2]^t_\eta=\sum^{{\rm min}\{s_1,t+1\}}_{r=
{\rm max}\{0,t+1-s_2\}}\lambda^{s_1s_2}_{tr}[v^r_{1\eta},v^{t+1-r}_{2\eta}]
\end{displaymath}
for some coefficients $\lambda^{s_1s_2}_{tr}$, where
$v_{1,2}\in\funcvecs{s_1,s_2}{\pi}{}$. The closed form identity gives the
relation:
\begin{displaymath}
(s_1+s_2-t)\lambda^{s_1s_2}_{(t-1)r}=(-)^{s_2+r+t+1}(s_1-r)
\lambda^{s_1s_2}_{t(r+1)}+(-)^{r+1}(s_2+r-t)\lambda^{s_1s_2}_{tr}
\end{displaymath}
which is solved to give:
\begin{displaymath}
[v_1,v_2]^t_\eta=\sum^{{\rm min}\{s_1,t+1\}}_{r=
{\rm max}\{0,t+1-s_2\}}(-)^{(s_2+t)(r+1)+s_1+1}[v^r_{1\eta},v^{t+1-r}_{2\eta}]
\end{displaymath}
The expansion is valid for $0\leq t\leq s_1+s_2-1$. In particular, in the $t=0$
case:
\begin{displaymath}
[v_1,v_2]_\eta=(-)^{s_1+1}\{\hat{v}_{1\eta}(v_{2\eta})+(-)^{s_2}\hat{v}_{2\eta}
(v_{1\eta})\}
\end{displaymath}
For the even functional multivector $\Gamma$,  $\Gamma_\eta$ is a bosonic local
functional, and the expression above gives:
\begin{displaymath}
\hat{\Gamma}_\eta(\Gamma_\eta)=-\frac{1}{2}[\Gamma,\Gamma]_\eta
\end{displaymath}
Thus $\Gamma_\eta$ is symmetric under $\hat{\Gamma}_\eta$ if and only if
$\Gamma$ is nilpotent at $\eta$, $[\Gamma,\Gamma]_\eta=0$.

When the bundle $\bundle{M}{B}{\pi}$ is a vector bundle, the fermionic
functional 1-form $\eta$ may be taken to be field independent,
$\eta=\int\eta_\alpha(x)\,d\phi^\alpha\wedge\Omega$ for a fermionic antifield
$\eta\in\Gamma(\bundle{M}{B^\ast}{})$. The functional multivector $v$ is
identified with the local functional $v_\eta=v[\phi,\eta]$ of the field $\phi$
and its antifield $\eta$, and the Schouten bracket induces the
Batalin-Vilkovisky antibracket, $[v_1,v_2]_\eta=(v_{1\eta},v_{2\eta})$, where:
\begin{displaymath}
(v_1,v_2)=\int\left\{\funcderv{v_1}{\eta_\alpha}{}\funcderv{v_2}{\phi}{\alpha}+
(-)^{s_1s_2}\funcderv{v_2}{\eta_\alpha}{}\funcderv{v_1}{\phi}{\alpha}\right\}
\Omega
\end{displaymath}
Note the symmetry in this bracket between $\phi$ and $\eta$, in particular the
role of the field and the antifield may be interchanged in the vector bundle
case. The commutator action of the functional multivector $v$ may be written as
an evolutionary vector field on $\bundle{M}{B\times B^\ast}{}$:
\begin{displaymath}
s_v=\int\Omega\left\{\funcderv{v}{\eta_\alpha}{}\parderv{}{}{\phi}{\alpha}{}+
(-)^s\funcderv{v}{\phi}{\alpha}\parderv{}{}{\eta}{}{\alpha}\right\}
\end{displaymath}

For the even functional multivector $\Gamma$, the BRST operator:
\begin{displaymath}
s_\Gamma=\int\Omega\left\{\funcderv{\Gamma}{\eta_\alpha}{}
\parderv{}{}{\phi}{\alpha}{}+\funcderv{\Gamma}{\phi}{\alpha}
\parderv{}{}{\eta}{}{\alpha}\right\}
\end{displaymath}
satisfies:
\begin{displaymath}
s^2_\Gamma=-\frac{1}{2}s_{[\Gamma,\Gamma]}
\end{displaymath}
and so $s_\Gamma$ is nilpotent if and only if $\Gamma$ is nilpotent, in which
case the $\Gamma$-cohomology on functional multivectors is identified with the
$s_\Gamma$-cohomology on local functionals of $\phi$ and $\eta$.

\section{The BRST-BV method in the jet bundle formalism}

\subsection{Symmetry in classical field theory}

The classical dynamics of a field $\phi\in\Gamma(\pi)$ is given, in the
Lagrangian formalism, by extremising the classical action
$S=\int{\cal L}\Omega\in\funcforms{0}{\pi}{}$ over $\Gamma(\pi)$. Thus the
classical field equation is:
\begin{displaymath}
\delta S[\phi]=0,\;{\rm or}\;E_\alpha({\cal L})\circ j^\infty\phi=0
\end{displaymath}
defining the extremal point of the action. Variations of the field that leave
the action invariant define infinitesimal symmetries of $S$. Thus the
evolutionary vector field $v$ is a symmetry of the classical action if $vS=0$.

The classical action may have a dependence on external, non-dynamical fields.
One example is the class of actions constructed by repeated contraction of the
even functional multivector $\Gamma=\sum^\infty_{r=0}\Gamma^{2r}$,
$\Gamma^{2r}\in\funcvecs{2r}{\pi}{}$. The action is then parametrized by the
space of fermionic functional 1-forms. The even functional multivector $\Gamma$
defines not only the action $\Gamma_\eta$, for the fermionic functional 1-form
$\eta$, but also the evolutionary vector field $\hat{\Gamma}_\eta$. A local
functional $J\in\funcforms{0}{\pi}{}$ is said to be symmetric under $\Gamma$ at
$\eta$ if $[\Gamma,J]_\eta=0$, in which case $\hat{\Gamma}_\eta(J)=0$. For any
closed fermionic functional 1-form
$\eta\in{\rm Ker}\,\delta\cap\funcforms{1}{\pi}{}$:
\begin{displaymath}
\hat{\Gamma}_\eta(\Gamma_\eta)=-\frac{1}{2}[\Gamma,\Gamma]_\eta
\end{displaymath}
so that $\hat{\Gamma}_\eta$ is a symmetry of the action $\Gamma_\eta$ when
$\Gamma$ satisfies the classical master equation at $\eta$:
\begin{displaymath}
[\Gamma,\Gamma]_\eta=0
\end{displaymath}
i.e.\ $\Gamma$ is nilpotent at $\eta$.

In practice, actions defined by even functional multivectors arise by
considering the extension of the original action $S$ to include terms defining
an algebra of symmetries of $S$. Suppose the action is symmetric under the even
functional multivector $v=\sum^\infty_{r=1}v^{2r}$,
$v^{2r}\in\funcvecs{2r}{\pi}{}$, i.e.\ $[v,S]=0$. The extension defined by $v$
is:
\begin{displaymath}
S_\eta=S+v_\eta=S+\sum^\infty_{r=1}\frac{1}{(2r)!}\,v^{2r}\intprod\eta
\stackrel{2r}{\intprod\cdots\intprod}\eta
\end{displaymath}
so that the extended action satisfies:
\begin{displaymath}
\hat{v}_\eta(S_\eta)=-\frac{1}{2}[v,v]_\eta
\end{displaymath}
More terms must be added to $v$, usually constructed from the algebra the
symmetries form, so that it satisfies the classical master equation, in which
case the extended action $S_\eta$ would be symmetric under $\hat{v}_\eta$.

\subsection{The functional integral}

The classical action determines the phases for a quantum field theory. The
Green's functions are given by:
\begin{displaymath}
\left\langle\,e^J\,\right\rangle=Z[J]/Z[0]
\end{displaymath}
where $J$ is a local functional and $Z[J]$ is the functional integral:
\begin{displaymath}
Z[J]=\int[d\phi]\,e^{\frac{i}{\hbar}S[\phi]+J[\phi]}
\end{displaymath}

The functional integral is a formal integral over the space $\Gamma(\pi)$ of
sections, and must be regularised before the measure $[d\phi]$ may be defined.
The regularisation is achieved by modifying the classical action $S[\phi]$ with
a regulator local functional $S_{\rm reg}[\phi;\Lambda]$, vanishing as the
regularisation is removed, $\Lambda\ra\infty$. The regularised functional
integral:
\begin{displaymath}
Z[0;\Lambda]=\int[d\phi]\,e^{\frac{i}{\hbar}(S[\phi]+
S_{\rm reg}[\phi;\Lambda])}
\end{displaymath}
is defined for each value of the regularisation parameter $\Lambda$. A
candidate for the partition function $Z[0]$ is given by the limit
$\lim_{\Lambda\ra\infty}Z[0;\Lambda]$. However this limit may be divergent, and
the limiting procedure must be accompanied by a renormalisation of the action
in order to achieve a finite result for $Z[0]$.

In the perturbation theory on linear bundles, the coefficient fields
${S_\alpha}^{I_1\ldots I_n}_{\beta_1\ldots\beta_n}$ appearing in the expansion
of the action:
\begin{displaymath}
S[\phi_0+\phi]=S[\phi_0]+\sum^\infty_{n=1}\frac{1}{(n+1)!}\int\phi^\alpha
\phi^{\beta_1}_{I_1}\ldots\phi^{\beta_n}_{I_n}{S_\alpha}^{I_1
\ldots I_n}_{\beta_1\ldots\beta_n}\Omega
\end{displaymath}
about a background solution $\phi_0$ of the classical field equation determine
the Feynman rules for the quantum field theory\footnote{For nonlinear bundles,
the bundle must be linearised before this analysis may be performed. This may
for instance be achieved by using geodesic coordinates, when the space $B$ is
endowed with a connection. For a background field $\phi_0$ the quantum field is
a vertical section $v$ of the tangent bundle to $B$ at $\phi_0$,
$v\in\Gamma(\bundle{M}{\phi^*_0VB}{})$. The vertical tangent field $v$ defines
geodesic deformations $e^v$ of the background field $\phi_0$, and the action
$S[\phi]$ defines a local functional $S[e^v\phi_0]$ on the bundle
$\bundle{M}{\phi^\ast_0VB}{}$. This functional may be used to generate the
Feynman rules.}. The requirement for regularisation manifests itself in the
need to regularise graphs with short distance (i.e.\ UV) divergences. The
coefficient fields are renormalised with $O(\hbar)$ counterterms, divergent as
the regularisation is removed, so that the limit
$\lim_{\Lambda\ra\infty}Z[0;\Lambda]$ is finite.

The renormalisation procedure must maintain the locality of the action,
i.e.\ the counterterms must not contain arbitrarily high order derivatives.
This cannot be guaranteed for the general local functional, as the counterterms
are dictated by the need to renormalise divergent graphs. Thus only certain
renormalisable local functionals may be taken as classical actions, determined
by the power counting rules.

To define the functional integral $Z[J]$, and hence define the Green's function
for the local functional operator $J$, extra counterterms must be added to
renormalise the graphs that include $J$-insertions. In order to make it clear
that the operator is to be understood only in the renormalised sense, use the
normal product notation $N_d(J)$, where $d$ is the UV dimension of $J$. The
ambiguity in the renormalisation procedure allows for finite renormalisations
of the normal product $N_d$---for a basis of local functional operators
$\{J_i\}$ with UV degree $d_i\leq d$, an alternative normal product is given
by:
\begin{displaymath}
\tilde{N}_d(J)=N_d\left(J+\sum_ic_iJ_i\right)
\end{displaymath}
with $O(\hbar)$ coefficients $c_i$. The renormalisation procedure frequently
causes operator expressions whose classical analogues vanish not to be zero,
but to be given by $O(\hbar)$ anomalous operators. The above arbitrariness in
the normal product may then be employed in an attempt to remove the anomalies.

Ward identities reflecting the presence of symmetries in the quantum field
theory may be generated from the formal Stokes' theorem for the functional
integral:
\begin{eqnarray*}
0&=&\int[d\phi]\,v\left(e^{\frac{i}{\hbar}S[\phi]+J[\phi]}\right)\\
&=&\int[d\phi]\,e^{\frac{i}{\hbar}S[\phi]+J[\phi]}\left(\frac{i}{\hbar}vS+vJ
\right)\![\phi]
\end{eqnarray*}
so that, if the action $S$ is invariant under the evolutionary vector field
$v$, $vS=0$, then the Ward identity for the  local functional $J$ is:
\begin{displaymath}
0=\left\langle\,vJ\,\right\rangle
\end{displaymath}
The renormalisation of the action and the symmetry cause extra anomalous
insertions in the above expression. In fact the validity of the Ward identity
relies on two conditions:
\begin{enumerate}
\item Invariance of the regulator: The regularised functional integral includes
the regulator term $S_{\rm reg}$. Thus at the regularised stage the expression
above should read:
\begin{displaymath}
0=\int[d\phi]\,e^{\frac{i}{\hbar}(S[\phi]+
S_{\rm reg}[\phi;\Lambda])+J[\phi]}\left(\frac{i}{\hbar}v(S+
S_{\rm reg})+vJ\right)\![\phi;\Lambda]
\end{displaymath}
resulting in an anomalous insertion after the limit $\Lambda\ra\infty$ is
taken, due to the renormalisation of the action when the regulator is not
invariant under the symmetry.
\item Invariance of the measure: It cannot be guaranteed that the regularised
measure $[d\phi]$ is invariant under the symmetry, implicitly assumed in the
application of the formal Stokes' theorem above. In general, the regularised
divergence $\Delta v$ is non-zero, and this results in an anomalous insertion
after the regularisation is removed, given by the renormalised divergence of
the symmetry. The renormalised divergence operator $\Delta$ defines a map from
the space of evolutionary vector fields $\funcvecs{1}{\pi}{}$ to the space of
local functionals $\funcforms{0}{\pi}{}$.
\end{enumerate}
The renormalised Ward identity is thus:
\begin{displaymath}
0=\left\langle\,vJ\,\right\rangle+\frac{i}{\hbar}\left\langle\,J\Delta\,
\right\rangle
\end{displaymath}
where the anomalous insertion $\Delta$ is due to the renormalisation of the
action and the divergence of the symmetry.

An even functional multivector $\Gamma$ may be functionally integrated after
repeated contraction with fermionic functional 1-forms, so that each fermionic
functional 1-form $\eta\in\funcforms{1}{\pi}{}$ defines the Green's functions
for a quantum field theory:
\begin{displaymath}
\left\langle\,e^J\,\right\rangle_\eta=Z_\eta[J]/Z_\eta[0]
\end{displaymath}
where $Z_\eta[J]$ is the functional integral:
\begin{displaymath}
Z_\eta[J]=\int[d\phi]\,e^{\frac{i}{\hbar}\Gamma_\eta[\phi]+J[\phi]}
\end{displaymath}
The definition above is dependent on the renormalisability of the action
$\Gamma_\eta$. In general, the quantum field theories are defined only on a
subset of the fermionic functional 1-forms, determined by the even functional
multivector action $\Gamma$.

Consider the dependence of the functional integral on the fermionic functional
1-form. Under a perturbation $\chi$ of $\eta$, the action $\Gamma_\eta$ changes
by $\hat{\Gamma}_\eta\intprod\chi$, to lowest order in $\chi$. Thus:
\begin{displaymath}
Z_{\eta+\chi}[J]-Z_\eta[J]=\frac{i}{\hbar}\int[d\phi]\,e^{\frac{i}{\hbar}
\Gamma_\eta[\phi]+J[\phi]}(\hat{\Gamma}_\eta\intprod\chi)[\phi]
\end{displaymath}
to lowest order in $\chi$. If the perturbation is exact, $\chi=\delta\psi$, the
perturbation of $\Gamma_\eta$ is $\hat{\Gamma}_\eta(\psi)$, so that
`integration by parts' may be performed in the functional integral:
\begin{displaymath}
Z_{\eta+\delta\psi}[J]-Z_\eta[J]=-\frac{i}{\hbar}\int[d\phi]\,
e^{\frac{i}{\hbar}\Gamma_\eta[\phi]+J[\phi]}\left(\frac{i}{\hbar}
\hat{\Gamma}_\eta(\Gamma_\eta)+\hat{\Gamma}_\eta(J)\right)\![\phi]\,\psi[\phi]
\end{displaymath}
The important case is when the fermionic functional 1-form is closed,
$\eta\in{\rm Ker}\,\delta\cap\funcforms{1}{\pi}{}$, so that:
\begin{displaymath}
Z_{\eta+\delta\psi}[J]-Z_\eta[J]=\frac{i}{\hbar}\int[d\phi]\,e^{\frac{i}{\hbar}
\Gamma_\eta[\phi]+J[\phi]}\left(\frac{i}{2\hbar}[\Gamma,\Gamma]_\eta+
[\Gamma,J]_\eta\right)\![\phi]\,\psi[\phi]
\end{displaymath}
explicitly displaying the relation between the functional integral and the
classical master equation. Thus if $\Gamma$ satisfies the classical master
equation, then the Green's functions for symmetric operators depend only on the
cohomology class of the closed fermionic functional 1-form $\eta$ in the
variational cohomology group $H_1(\pi,\delta)$. This result is important for
two reasons, explicitly demonstrated in later sections. Firstly, this
expression generates the Lam action principle for the quantum effective action,
and hence generates the anomaly consistency condition. Secondly, when the
action exhibits gauge symmetries, and the propagators are undefined, the `gauge
fixed' extension $\Gamma_\eta$ may be used instead of the original action. The
above expression gives the dependence of the resulting quantum field theory on
the fermionic gauge fixing functional 1-form $\eta$.

The validity of the above formal manipulation of the functional integral
requires a renormalisation procedure that does not introduce anomalous
insertions, i.e.\ the regulator and measure must be invariant:
\begin{enumerate}
\item Invariance of the regulator: The functional integral is regularised with
an even functional multivector regulator $\Gamma_{\rm reg}$ (this allows for
actions $\Gamma_\eta$ defined by different fermionic functional 1-forms $\eta$
to be regularised with different regulators $\Gamma_{{\rm reg}\,\eta}$). Thus
even when the action $\Gamma$ satisfies the classical master equation, the
regularised action $\Gamma+\Gamma_{\rm reg}$ may not, and residual anomalous
odd functional multivector insertions may remain after the regularisation is
removed.
\item Invariance of the measure: The formal manipulations in the regularised
functional integral must include divergence terms when the measure is not
invariant under the evolutionary vector field $\hat{\Gamma}_\eta$. The
renormalised divergence operator $\Delta$ on evolutionary vector fields defines
a divergence operator $\Delta$ on the functional multivectors, given by:
\begin{displaymath}
(\Delta v)_\eta=\Delta(\hat{v}_\eta)
\end{displaymath}
The divergence operator is effectively a coderivative to the variational
derivative, mapping down the functional multivectors:
\begin{displaymath}
\begin{array}{cccccccccc}
\makebox[13mm]{$\cdots$} &
\makebox[0mm]{$\funcvecs{s+1}{\pi}{}$} &
\makebox[13mm]{$\stackrel{\Delta}{\ra}$} &
\makebox[0mm]{$\funcvecs{s}{\pi}{}$} &
\makebox[13mm]{$\cdots$} &
\makebox[0mm]{$\funcvecs{1}{\pi}{}$} &
\makebox[13mm]{$\stackrel{\Delta}{\ra}$} &
\makebox[0mm]{$\funcvecs{0}{\pi}{}$} &
\makebox[13mm]{$\ra$} &
\makebox[0mm]{$0$} \\
\end{array}
\end{displaymath}
given by the renormalised trace $\Delta v={\rm tr}(\delta v)$. Note however
that $\Delta$ is not necessarily nilpotent, in general $\Delta^2$ will be
proportional to the renormalised Ricci tensor on the space of sections
$\Gamma(\pi)$.
\end{enumerate}
Both these factors result in a possible anomalous odd functional multivector
insertion $\Delta$ in the formal expressions above, and may destroy the
independence of the Green's functions on the fermionic functional 1-form.

The even functional multivector $v=\sum^\infty_{r=1}v^{2r}$ defines extensions
to the classical action, and each extended action determines the phases for a
quantum field theory. Proceeding as before, consider the functional integral:
\begin{displaymath}
Z_\eta[J]=\int[d\phi]\,e^{\frac{i}{\hbar}S_\eta[\phi]+J[\phi]}
\end{displaymath}
related to the Green's functions via:
\begin{displaymath}
Z_\eta[J]=Z_\eta[0]\left\langle\,e^J\,\right\rangle_\eta
\end{displaymath}
If $\eta$ is a closed fermionic functional 1-form, and $\psi$ is a fermionic
local functional, then to lowest order in $\psi$:
\begin{displaymath}
Z_{\eta+\delta\psi}[J]-Z_\eta[J]=\frac{i}{\hbar}\int[d\phi]\,e^{\frac{i}{\hbar}
S_\eta[\phi]+J[\phi]}\left(\frac{i}{2\hbar}[v,v]_\eta+[v,J]_\eta\right)\![\phi]
\,\psi[\phi]
\end{displaymath}
Thus if the even functional multivector $v$ can be modified to solve the
classical master equation at $\eta$, $[v,v]_\eta=0$, then the Green's functions
for functionals symmetric under $v$ at $\eta$ are unchanged by exact
perturbations of $\eta$. As before, renormalisation introduces extra anomalous
terms into the above expression, due to regulators for both the action $S$ and
the even functional multivector $v$, and the possible non-vanishing divergence
of $v$.

The analysis above can be repeated in the field-antifield formalism when the
bundle $\bundle{M}{B}{\pi}$ is a vector bundle (more generally, the analysis
applies locally for a nonlinear bundle when it is linearised). Take the
fermionic functional 1-form to be field independent,
$\eta=\int\eta_\alpha(x)\,d\phi^\alpha\wedge\Omega$ for a fermionic antifield
$\eta\in\Gamma(\bundle{M}{B^\ast}{})$. Functional multivectors are then
identified with their contracted counterparts, local functionals of the field
$\phi$ and its antifield $\eta$.

Define the generating functional for the Green's functions:
\begin{displaymath}
Z[j]=\int[d\phi]\,e^{\frac{i}{\hbar}S[\phi]+\int j_\alpha\phi^\alpha\Omega}
\end{displaymath}
Ward identities corresponding to the symmetry of the action $S$ under the
evolutionary vector field $v$ are generated from the formal functional integral
expression:
\begin{displaymath}
0=\int[d\phi]\,e^{\frac{i}{\hbar}S[\phi]+
\int j_\alpha\phi^\alpha\Omega}\left(\int j_\alpha v^\alpha\Omega\right)
\end{displaymath}
so that the Ward identity for the $n$-point Green's function is:
\begin{displaymath}
0=\frac{i}{\hbar}\left\langle\,\phi^{\alpha_1}(x_1)\ldots\phi^{\alpha_n}(x_n)\,
\Delta\,\right\rangle+\sum^n_{r=1}\left\langle\,\phi^{\alpha_1}(x_1)
\ldots v^{\alpha_r}(x_r)\ldots\phi^{\alpha_n}(x_n)\,\right\rangle
\end{displaymath}
where the presence of the anomalous insertion $\Delta$ reflects the possible
lack of invariance of the measure and the regulator under the symmetry.

For the even functional multivector action $\Gamma$, construct the generating
functional in the presence of the external antifield:
\begin{displaymath}
Z[j,\eta]=\int[d\phi]\,e^{\frac{i}{\hbar}\Gamma[\phi,\eta]+
\int j_\alpha\phi^\alpha\Omega}
\end{displaymath}
The dependence on the antifield may be analysed:
\begin{eqnarray*}
\int j_\alpha\funcderv{Z}{\eta_\alpha}{}\Omega&=&\frac{i}{\hbar}\int[d\phi]\,
e^{\frac{i}{\hbar}\Gamma[\phi,\eta]}\left\{\int\Omega
\funcderv{\Gamma}{\eta_\alpha}{}\parderv{}{}{\phi}{\alpha}{}\right\}
\left(e^{\int j_\alpha\phi^\alpha\Omega}\right)\\
&=&\frac{1}{\hbar^2}\int[d\phi]\,e^{\frac{i}{\hbar}\Gamma[\phi,\eta]+
\int j_\alpha\phi^\alpha\Omega}\left(\int\funcderv{\Gamma}{\eta_\alpha}{}
\funcderv{\Gamma}{\phi}{\alpha}\Omega\right)
\end{eqnarray*}
If the even functional multivector $\Gamma$ satisfies the classical master
equation, then the generating functional satisfies the Lam action principle:
\begin{displaymath}
\int j_\alpha\funcderv{Z}{\eta_\alpha}{}\Omega=0
\end{displaymath}
The introduction of an even functional multivector regulator $\Gamma_{\rm reg}$
and the lack of invariance of the measure introduce extra anomalous insertions
into the above expression. The renormalised Lam action principle for the
generating functional is thus:
\begin{displaymath}
\int j_\alpha\funcderv{Z}{\eta_\alpha}{}\Omega=\frac{1}{\hbar^2}\Delta\circ Z
\end{displaymath}
where $\Delta$ is an odd functional multivector insertion.

The quantum effective action $\Gamma_q$ is constructed by taking the Legendre
transform of the (connected) generating functional with respect to the source
field $j$, and is a non-local even functional multivector. By taking the
Legendre transform of the Lam action principle for the generating functional,
the Lam action principle for the quantum effective action is seen to be:
\begin{displaymath}
[\Gamma_q,\Gamma_q]=\Delta\circ\Gamma_q
\end{displaymath}
For a general bundle $\bundle{M}{B}{\pi}$, the quantum effective action
$\Gamma_q$ at the section $\phi$ may be constructed from a linear approximation
to the bundle about $\phi$. With this definition of the quantum effective
action, the Lam action principle above remains valid for nonlinear bundles.

\subsection{Anomalies and the consistency condition}

Consider the consequences of the Lam action principle:
\begin{displaymath}
[\Gamma_q,\Gamma_q]=\Delta\circ\Gamma_q
\end{displaymath}
for the anomalous odd functional multivector insertion $\Delta$. Take the
commutator with $\Gamma_q$---the Jacobi identity for the Schouten bracket
gives:
\begin{displaymath}
[\Gamma_q,\Delta\circ\Gamma_q]=0
\end{displaymath}

It is observed in the perturbation theory that the quantum effective action may
be constructed by summing over the 1-particle irreducible graphs. Thus in
particular, in the $\hbar$-expansion of $\Gamma_q$:
\begin{displaymath}
\Gamma_q=\sum^\infty_{r=0}\hbar^r\Gamma_{q\,r}
\end{displaymath}
the first term is given by the classical action, $\Gamma_{q\,0}=\Gamma$.
Furthermore, if the anomalous insertion is expanded as
$\Delta=\hbar^r\Delta_r+O(\hbar^{r+1})$, then:
\begin{displaymath}
\Delta\circ\Gamma_q=\hbar^r\Delta_r+O(\hbar^{r+1})
\end{displaymath}
The lowest order term in the $\hbar$-expansion of the expression
$[\Gamma_q,\Delta\circ\Gamma_q]=0$ is:
\begin{displaymath}
[\Gamma,\Delta_r]=0
\end{displaymath}
This is the consistency condition for the anomaly---the lowest order anomaly
term must be symmetric under $\Gamma$.

The case of interest is when the classical action $S$ is extended by an even
functional multivector $v=\sum^\infty_{r=1}v^{2r}$ satisfying $[v,S]=0$, and in
particular when $v$ may be taken to be a solution of the classical master
equation $[v,v]=0$. The lowest order anomaly must thus be $\Gamma$-closed,
$[\Gamma,\Delta_r]=0$, where $\Gamma=S+v$. Suppose the anomaly is
$\Gamma$-exact, $\Delta_r=[\Gamma,\Gamma_r]$ for some even functional
multivector $\Gamma_r$. A finite renormalisation of the extended action
$\Gamma$ can be used to remove the anomaly:
\begin{displaymath}
\Gamma\ra\Gamma-\frac{1}{2}\hbar^r\Gamma_r
\end{displaymath}
so that $[\Gamma_q,\Gamma_q]=O(\hbar^{r+1})$. Thus the anomaly is non-trivial
if and only if it defines a non-trivial cohomology class in the odd
$\Gamma$-cohomology group $H^{\rm odd}(\pi,\Gamma)$.

\subsection{Gauge fixing}

As an example of the use of the jet formalism, consider the case of a classical
action with gauge symmetries. The gauge symmetries define degeneracies in the
equations of motion, and prevent the propagators from being defined. The action
must be gauge fixed to remove the degeneracies. An obvious possibility is to
use an even functional multivector extension, and consider the dependence of
the resulting quantum field theories on the gauge fixing fermionic functional
1-form.

The gauge symmetries are parametrised by fields from a vector bundle,
necessarily the collection of fields must be extended to include the parameter
field. However, the analysis proceeds as before without requiring quantisation
of the parameter field. In fact, in order to generate the Faddeev-Popov
determinant in the functional integral, it is the antifield to the parameter
field, the ghost field, that must be quantised. This does not invalidate the
analysis of previous sections for this case, as the symmetry between the field
and its antifield in the Batalin-Vilkovisky antibracket may be exploited.
Effectively, the ghost field is treated as the field, with the parameter field
appearing as the antifield to the ghost field.

In addition to the ghost field, there is the antifield to the $\phi$-field.
However, in the general nonlinear bundle case the antifield cannot be defined.
The solution to this is to work with the gauge fixing fermion in place
throughout the calculation. The gauge fixing fermion thus appears naturally as
the $\phi$-component of the fermionic functional 1-form.

Consider the classical action $S$ with gauge symmetries:
\begin{displaymath}
v^{R\alpha}\funcderv{S}{\phi}{\alpha}=0
\end{displaymath}
for some coefficient operators $v^{R\alpha}=v^{R\alpha I}d_I$. The transitions
for the $R$-index of $v^{R\alpha}$ determine a bundle $\bundle{M}{E}{\tau}$,
with adapted local coordinates $(x^i,b^R)$ on $E$\footnote{For example, in the
case of the Yang-Mills theory based on a principal fibre bundle
$\bundle{M}{P}{}$ with Lie group $G$, the bundle $\bundle{M}{E}{\tau}$ is the
associated bundle with fibre the Lie algebra $g$ of $G$ in the adjoint
representation, and the gauge symmetry operator $v^{R\alpha}$ is given by the
covariant derivative.}. The gauge symmetry expression may be written in
functional multivector notation as  $[v,S]=0$, where $v$ is the functional
bivector:
\begin{displaymath}
v=\int\Omega\parderv{}{}{b}{S}{}\wedge v^{S\alpha}\parderv{}{}{\phi}{\alpha}{}
\end{displaymath}
in $\funcvecs{2}{(\pi\times\tau)}{}$. Thus $v\intprod\omega$ is a symmetry of
the action for any functional 1-form
$\omega\in\funcforms{1}{(\pi\times\tau)}{}$.

Assume the bundle $\bundle{M}{E}{\tau}$ is a vector bundle. Restrict attention
to fermionic functional 1-forms $\eta$ of the form:
\begin{displaymath}
\eta=\int c_Sdb^S\wedge\Omega+\delta\Psi
\end{displaymath}
for the fermionic ghost field $c\in\Gamma(\bundle{M}{E^\ast}{})$ and the gauge
fixing fermion $\Psi\in\funcforms{0}{\pi}{}$. The action $S$ may be extended by
the contracted functional bivector $v_\eta$, giving an action dependent on the
ghost field as well as the $\phi$-field (but independent of the $b$-field):
\begin{displaymath}
S_\Psi[\phi,c]=S[\phi]+\int c_Sv^{S\alpha}\funcderv{\Psi}{\phi}{\alpha}\Omega
\end{displaymath}

Consider the functional integral:
\begin{displaymath}
Z_\Psi[J;c]=\int[d\phi]\,e^{\frac{i}{\hbar}S_\Psi[\phi,c]+J[\phi]}
\end{displaymath}
for the local functional $J\in\funcforms{0}{\pi}{}$. Analyse the dependence of
$Z_\Psi[J;c]$ on the gauge fixing fermion $\Psi$. For the fermionic local
functional $\tilde{\Psi}\in\funcforms{0}{\pi}{}$:
\begin{eqnarray*}
Z_{\Psi+\tilde{\Psi}}[J;c]-Z_\Psi[J;c]&=&\frac{i}{\hbar}\int[d\phi]\,
e^{\frac{i}{\hbar}S_\Psi[\phi,c]+J[\phi]}\\
&&\times\left(\frac{i}{2\hbar}[v,v]_\eta+[v,J]_\eta\right)\![\phi,c]\,
\tilde{\Psi}[\phi]
\end{eqnarray*}
to lowest order in $\tilde{\Psi}$ (ignoring anomalous terms). Care must be
taken when proving this statement. Since functional integration over the
$b$-field is not included in the functional integral, the above expression does
not follow directly from the expressions of previous sections. In fact the lack
of functional integration over the $b$-field does not introduce extra terms
since the ghost field, the gauge fixing fermion, and the extended action are
all independent of the $b$-field.

The extended action determines a gauge fixed quantum field theory. Take the
gauge fixing fermion $\Psi$ in the form:
\begin{displaymath}
\Psi=\int f_S(\phi)\bar{c}^S\Omega
\end{displaymath}
for the fermionic antighost field $\bar{c}\in\Gamma(\bundle{M}{E}{\tau})$ and
the gauge fixing functions $f_S$. If the ghost and antighost fields are
quantised, i.e.\ the functional integral is taken to be:
\begin{displaymath}
Z_f[J]=\int[d\phi][dc][d\bar{c}]\,e^{\frac{i}{\hbar}S_\Psi[\phi,c,\bar{c}]+
J[\phi]}
\end{displaymath}
then the functional integral formally contains the Faddeev-Popov determinant:
\begin{eqnarray*}
Z_f[J]&=&\int[d\phi]\,e^{\frac{i}{\hbar}S[\phi]+J[\phi]}\left(
\int[dc][d\bar{c}]\,e^{\frac{i}{\hbar}
\int c_Rv^{R\alpha}E_\alpha(f_T\bar{c}^T)\Omega}\right)\\
&=&\int[d\phi]\,\delta(f_S(\phi))e^{\frac{i}{\hbar}S[\phi]+J[\phi]}
\end{eqnarray*}
corresponding to the gauge-fixing conditions $f_S=0$.

For $h,g\in\funcforms{0}{(\pi\times\tau)}{}$, the commutator
$[v\intprod\delta h,v\intprod\delta g]$ is also a symmetry of the action.
Suppose the algebra of symmetries generated by
$\funcforms{0}{(\pi\times\tau)}{}$ closes on
$\funcforms{0}{(\pi\times\tau)}{}$. This defines a Poisson bracket on
$\funcforms{0}{(\pi\times\tau)}{}$:
\begin{displaymath}
[v\intprod\delta h,v\intprod\delta g]=v\intprod\delta\{h,g\}
\end{displaymath}
given by a nilpotent functional bivector $\Gamma$ via
$\{h,g\}=\Gamma\intprod\delta h\intprod\delta g$. By counting the powers of the
fields in the defining expression above, the algebra functional bivector
$\Gamma$ must take the form:
\begin{displaymath}
\Gamma=\frac{1}{2}\int\Omega\parderv{}{}{b}{R}{}
\wedge b^S_I{D^{RT}}^I_S\parderv{}{}{b}{T}{}
\end{displaymath}
for some $b$-independent operator ${D^{RT}}^I_S$ (necessarily dependent on the
$\phi$-field when the bundle $\bundle{M}{B}{\pi}$ is nonlinear\footnote{This is
a significant generalisation of the Yang-Mills case, where the algebra is given
by the functional bivector:
\begin{displaymath}
\Gamma=\frac{1}{2}\int\Omega\parderv{}{}{b}{R}{}
\wedge b^Sf^{RT}_S\parderv{}{}{b}{T}{}
\end{displaymath}
where the coefficients $f^{RT}_S$ are the structure constants of the Lie
algebra $g$. Thus in the Yang-Mills case the operator ${D^{RT}}^I_S$ is a
constant, field independent matrix operator, vanishing for $I\neq0$.}).

Consider the consequences of the closure of the algebra for the contracted
functional multivectors. The algebra may be written in terms of the
evolutionary vector fields $\hat{v}_\eta$ as:
\begin{displaymath}
[\hat{v}_\eta,\hat{v}_\eta]=2[v,\Gamma_\eta]
\end{displaymath}
This expression is contracted with $\eta$ to get a local functional expression:
\begin{displaymath}
[v,v]_\eta=\hat{v}_\eta(\Gamma_\eta)
\end{displaymath}
Combine this with the expression $[\Gamma,\Gamma]_\eta=0$ for the nilpotency of
$\Gamma$ to obtain:
\begin{displaymath}
[v+\frac{1}{2}\Gamma,v+\frac{1}{2}\Gamma]_\eta+\hat{\Gamma}_\eta(v_\eta)=0
\end{displaymath}
The functional bivector $v+\frac{1}{2}\Gamma$ is thus nilpotent on fermionic
functional 1-forms for which $\hat{\Gamma}_\eta(v_\eta)$ vanishes. In
particular, for fermionic functional 1-forms of the form
$\eta=\int c_Sdb^S\wedge\Omega+\delta\Psi$ the local functional $v_\eta$ is
independent of the $b$-field. Thus $\hat{\Gamma}_\eta(v_\eta)=0$, and hence
$v+\frac{1}{2}\Gamma$ satisfies the classical master equation at $\eta$.

In the case of a closed algebra, include a term dependent on the external
$b$-field in the extended action. For
$\eta=\int c_Sdb^S\wedge\Omega+\delta\Psi$, the extended action becomes:
\begin{eqnarray*}
S_\Psi[\phi,c;b]&=&S[\phi]+v_\eta[\phi,c]+\frac{1}{2}\Gamma_\eta[\phi,c;b]\\
&=&S[\phi]+\int c_Sv^{S\alpha}\funcderv{\Psi}{\phi}{\alpha}\Omega+
\frac{1}{4}\int c_Rb^S_I{D^{RT}}^I_Sc_T\Omega
\end{eqnarray*}
Consider the functional integral:
\begin{displaymath}
Z_\Psi[J;b]=\int[d\phi][dc]\,e^{\frac{i}{\hbar}S_\Psi[\phi,c;b]+J[\phi]}
\end{displaymath}
obtained by quantising the ghost field, leaving the $b$-field as an external
field (this reversal of the roles of the $b$- and $c$-fields is necessary since
the extended action is linear in the $b$-field, and hence there are
difficulties in defining the functional integration over the $b$-field). The
symmetry between the $b$- and $c$-fields in the Batalin-Vilkovisky antibracket
is now exploited to generate the expression for the dependence of the
functional integral on the gauge fixing fermion. For
$\eta=\int c_Sdb^S\wedge\Omega+\delta\Psi$, the expression above for the
bracket $[v+\frac{1}{2}\Gamma,v+\frac{1}{2}\Gamma]_\eta$ is written out in
full:
\begin{eqnarray*}
0&=&\left(-2\int\Omega\,c_Sv^{S\alpha}\parderv{}{}{\phi}{\alpha}{}\right)
(v_\eta+\frac{1}{2}\Gamma_\eta)+2\int
\funcderv{(v_\eta+\frac{1}{2}\Gamma_\eta)}{c_S}{}
\funcderv{(v_\eta+\frac{1}{2}\Gamma_\eta)}{b}{S}\Omega\\
&=&\int\Omega\,\left\{-2c_Sv^{S\alpha}\parderv{}{}{\phi}{\alpha}{}+\frac{1}{2}
(c_R{D^{RT}}^I_Sc_T)d_I\parderv{}{}{c}{}{S}\right\}(v_\eta+\frac{1}{2}
\Gamma_\eta)
\end{eqnarray*}
The above expression\footnote{This expression contains the evolutionary vector
field:
\begin{displaymath}
\int\Omega\,\left\{c_Sv^{S\alpha}\parderv{}{}{\phi}{\alpha}{}-
\frac{1}{4}(c_R{D^{RT}}^I_Sc_T)d_I\parderv{}{}{c}{}{S}\right\}
\end{displaymath}
corresponding to the BRST variations of the $\phi$- and $c$-fields in the
Yang-Mills theory.} can now be placed under the functional integral sign, and
manipulating the functional integral as before gives:
\begin{displaymath}
Z_{\Psi+\tilde{\Psi}}[J;b]-Z_\Psi[J;b]=\frac{i}{\hbar}\int[d\phi][dc]\,
e^{\frac{i}{\hbar}S_\Psi[\phi,c;b]+J[\phi]}[v,J]_\eta[\phi,c]\,\tilde{\Psi}
[\phi]
\end{displaymath}
to lowest order in $\tilde{\Psi}$ (neglecting anomaly terms). It should be
clear from this example that the formal expressions of the previous sections
have analogous expressions when the roles of the fields and the antifields are
reversed. The functional integral, where defined, is thus independent of the
gauge fixing fermion for symmetric sources $J$ (in the absence of anomaly
terms). In particular, if the physical operators are required to be symmetric,
then the quantum field theories constructed from different gauge fixing
fermions are equivalent.

\pagebreak

\section{Acknowledgements}

Thanks to Paul Howe for his encouragement with this article. This research is
funded by the Science and Engineering Research Council.

\end{document}